\documentclass[twoside]{article}
\usepackage{url}
\usepackage{qic}

\renewcommand{\proof}[1]{{\bf Proof.} #1\hfill \square}
\newtheorem{conjecture}{Conjecture}
\newtheorem{proposition}{Proposition}

\usepackage{amstext,amsmath}

\DeclareSymbolFont{AMSb}{U}{msb}{m}{n}
\DeclareSymbolFontAlphabet{\mathbb}{AMSb}
\def\C{{\mathbb{C}}}
\def\K{{\mathbb{K}}}
\def\Q{{\mathbb{Q}}}
\def\Z{{\mathbb{Z}}}

\def\DFT{{\rm DFT}}
\def\bra#1{\langle#1|}
\def\ket#1{|#1\rangle}
\def\ip#1#2{\langle#1|#2\rangle}

\def\Tr{\mathop{{\rm Tr}}}
\def\entspricht{\mathrel{\widehat{=}}}

\def\Magma{{\sf MAGMA}}
\def\MUB{\mathcal{MUB}}

\begin{document}
\setlength{\textheight}{8.0truein}    

\def\mykeywords{Mutually unbiased bases, SIC-POVMs, finite geometry}

\def\mytitle{On SIC-POVMs and MUBs in Dimension~6}
\runninghead{\mytitle}
            {M. Grassl}

\setcounter{page}{1}

\normalsize\textlineskip
\thispagestyle{empty}
\setcounter{page}{1}

\alphfootnote

\fpage{1}

\vspace*{0.035truein}
\centerline{\bf\uppercase{\mytitle}}
\vspace*{0.037truein}
\centerline{\footnotesize
MARKUS GRASSL}
\centerline{\footnotesize\it IAKS, Arbeitsgruppe Prof. Dr. Thomas Beth, 
Universit\"at Karlsruhe}
\baselineskip=10pt
\centerline{\footnotesize\it Am Fasanengarten 5, 76128 Karlsruhe, Germany}
\centerline{\footnotesize\it E-mail: {\sf grassl@ira.uka.de}}

\centerline{\footnotesize June 23, 2004, updated May 25, 2009}
\vspace*{10pt}

\abstracts{We provide a partial solution to the problem of
    constructing mutually unbiased bases (MUBs) and symmetric
    informationally complete POVMs (SIC-POVMs) in non-prime-power
    dimensions.}{An algebraic description of a SIC-POVM in dimension
    six is given.  Furthermore it is shown that several sets of three mutually
    unbiased bases in dimension six are maximal, i.e., cannot be extended.}{}

\vspace*{10pt}

\keywords{\mykeywords}

\vspace*{1pt}\textlineskip    
\section{Introduction}
In a recent paper, Wootters discusses relations between quantum
measurements and finite geometries \cite{Woo04:fin_geom}.  Both
projective quantum measurements corresponding to so-called mutually
unbiased bases (MUBs) and generalized quantum measurements described
by symmetric informationally complete positive operator valued
measures (SIC-POVMs) can be linked to finite affine planes.  While
some constructions of MUBs are directly related to affine planes over
finite fields (see, e.g., \cite{GHW04}), the link between SIC-POVMs
and finite affine planes is on a more abstract level.  Both finite
affine planes and maximal sets of MUBs can be constructed if the
dimension is a prime power (see, e.g., \cite{BBRV02,KlRo04}). In
neither case constructions for non-prime-powers are known, but there
is numerical evidence that SIC-POVMs exist in all dimensions
\cite{RBKSC04}.

In this note we provide an explicit algebraic expression for a SIC-POVM in
dimension six.  This shows that the problem of constructing SIC-POVMs
is not equivalent to the existence of a finite affine plane, as it is
well-known that there is no finite affine plane of order six
\cite{Tarry1900}.  For MUBs in dimension six, however, the question
remains open whether a maximal set of seven MUBs exists.  It is known
that three MUBs in dimension six exist, and it is widely believed that
there are no four MUBs.  Here we show that particular sets of three
MUBs are maximal in the sense that they cannot be extended to four
MUBs.

\section{The Weyl-Heisenberg Group and the Jacobi Group}
Before we discuss the construction of SIC-POVMs and MUBs, we recall
properties of the well-known Weyl-Heisenberg group and investigate its
automorphism group.

The Heisenberg group $H_d$ is a finite subgroup of $GL(d,\C)$ generated by
the cyclic shift operator $X$ and the phase operator $Z$ given by
\begin{equation}
X:=\sum_{j=0}^{n-1}\ket{j+1}\bra{j}
\qquad\mbox{and}\qquad
Z:=\sum_{j=0}^{n-1}\omega_d^j\ket{j}\bra{j},\label{eq:Heisenberg}
\end{equation}
where $\omega_d:=\exp(2\pi i/d)$ is a primitive $d$-th root of unity
and the cyclic shift is modulo $d$.  The group $H_d$ is also called
Weyl-Heisenberg group as the operators $X$ and $Z$ are the discrete
Weyl operators (see \cite[Kap.~IV.D]{Weyl28} and
\cite[Chapt.~IV.D]{Weyl50}).  Each element of $H_d$ can be uniquely
written as $\omega_d^c X^a Z^b$ with $a,b,c\in\{0,\ldots,d-1\}$.  Two
elements $\omega_d^c X^a Z^b$ and $\omega_d^{c'} X^{a'} Z^{b'}$ commute iff
$ab'-a'b=0 \bmod d$.  

The center $\zeta(H_d)$ of the group $H_d$ is generated by $\omega_d
I$, where $I$ denotes the identity matrix.  Ignoring the global phase
factor, the group $H_d$ is isomorphic to the direct product of two
cyclic groups of order $d$, i.e., $H_d/\zeta(H_d)\cong\Z_d\times \Z_d$
where $\Z_d:=\Z/d\Z$ denotes the ring of integers modulo $d$.  The
matrices $X^a Z^b$ are mutually orthogonal with respect to the trace
inner product and form a vector space basis of all $d\times d$
matrices.

In our constructions we also consider the normalizer of the Heisenberg
group $H_d$ in the group $U(d,\C)$ of unitary matrices.  The Fourier
matrix
\begin{equation}
\DFT_d:=\frac{1}{\sqrt{d}}\bigl(\omega_d^{jk}\bigr)_{j,k=0}^{d-1}
\end{equation}
acts via conjugation on $H_d$.  It interchanges the operators $X$ and $Z$
(up to a phase factor).  Another matrix that acts on $H_d$ modulo the
center is the matrix $P_d$ which is defined as
\begin{displaymath}
P_d^{\text{even}}:=\sum_{j=0}^{d-1}\omega_d^{j^2/2}\ket{j}\bra{j},
\qquad\mbox{if $d$ is even,}
\end{displaymath}
and
\begin{displaymath}
P_d^{\text{odd}}:=\sum_{j=0}^{d-1}\omega_d^{j(j-1)/2}\ket{j}\bra{j},
\qquad\mbox{if $d$ is odd.}
\end{displaymath}
If $d$ is odd, $P_d^{\text{odd}}$ acts on $H_d$ via conjugation.  If
$d$ is even, conjugation with $P_d^{\text{even}}$ might introduce an
additional overall phase factor $\omega_d^{1/2}=\omega_{2d}$.
Therefore we enlarge the center of $H_d$ by the scalar multiple of
identity $\omega_{2d}I$. Then $P_d$ acts on $\langle
H_d,\omega_{2d}I\rangle$ via conjugation. What is more, the group
generated by $\DFT_d$ and $P_d$ acts transitively on $H_d$ modulo the
center.  We define the {\em Jacobi group} as the group which is generated
by $X$, $Z$, $\DFT_d$, and $P_d$, i.e.,
\begin{displaymath}
J_d:=\langle X, Z, \DFT_d, P_d\rangle.
\end{displaymath}
The Jacobi group is widely studied in the context of number theory
\cite{BeSc98}.  In quantum computing, the group is also known as the
Clifford group.  We essentially use the fact that the Jacobi group is
a semi-direct product of the Heisenberg group $H_d$ and $SL(2,\Z_d)$.
The Heisenberg group is a normal subgroup of the Jacobi group, and the
quotient group $J_d/\langle \zeta(J_d),H_d\rangle$ is isomorphic to
$SL(2,\Z_d)$.

\section{Symmetric Informationally Complete POVMs}
The most general quantum measurement is represented by positive
operator valued measures (POVMs).  A POVM is called {\em
informationally complete} if the statistics of the measurement allows
the reconstruction of the quantum state on which the measurement is
carried out.  As a quantum state in dimension $d$ is described by
$d^2-1$ real parameters, an informationally complete POVM must have at
least $d^2$ elements.  In order to achieve maximal statistical
independence of the outcomes, the inner product between the elements
should be constant.  Such a POVM is called a {\em symmetric,
informationally complete} POVM, or SIC-POVM for short.  It consists of
$d^2$ operators of the form $E_j=\Pi_j/d$, where the rank-one
projectors $\Pi_j=\ket{\phi_j}\bra{\phi_j}$ satisfy the condition
\begin{equation}\label{eq:SICPOVM:proj}
\Tr \Pi_j \Pi_k =\frac{1}{d+1}\qquad\mbox{for $j\ne k$,}
\end{equation}
or, equivalently,
\begin{equation}\label{eq:SICPOVM:state}
|\ip{\phi_j}{\phi_k}|^2 =\frac{1}{d+1}\qquad\mbox{for $j\ne k$.}
\end{equation}
It has been conjectured that SIC-POVMs exist in all dimensions
\cite{Zau99,RBKSC04}.
\begin{conjecture}[Zauner~\cite{Zau99}]
For every dimension $d\ge 2$ there exists a SIC-POVM whose elements
are the orbit of a positive rank-one operator $E_0$ under the
Heisenberg group.  What is more, $E_0$ commutes with an element $T$ of
the Jacobi group $J_d$. The action of $T$ on $H_d$ modulo the center
has order three.
\end{conjecture}
Zauner provides algebraic solutions for dimensions $d=2,3,4,5$, and
numerical solutions for $d=6,7$.  A weaker form of his
conjecture---the existence of SIC-POVMs that are the orbit under the
Heisenberg group---has been numerically verified up to dimension
$d=45$ \cite{RBKSC04} (as of May 25, 2009, Andrew Scott has found
numerical solutions up to $d=67$ \cite{Sco09}).

A SIC-POVM that is invariant under the Heisenberg group is given by
\begin{equation}
\{ h E_0 h^{-1}\colon h\in H_d\},
\end{equation}
where $E_0=\ket{\phi_0}\bra{\phi_0}/d$ for some normalized quantum
stated $\ket{\phi_0}\in\C^d$.  The image of the corresponding
projectors $\Pi_i$ is the set
\begin{equation}\label{eq:Heisenberg_orbit}
\{ X^a Z^b \ket{\phi_0}\colon a,b=0,\ldots,d-1\}.
\end{equation}
Note that a global change of basis $U$ yields another SIC-POVM which
is invariant under the conjugated group $U H_d U^{-1}$.  If $U$ is an
element of the Jacobi group, we get an equivalent SIC-POVM that is
invariant under the very same representation of the Heisenberg group.

In the following we briefly sketch how we found an algebraic solution
for a SIC-POVM in dimension $d=6$.  Starting from Zauner's conjecture,
we search for an initial vector $\ket{\phi_0}$ that is an eigenvector
of an element of the Jacobi group.  The Jacobi group $J_6$ has 124416
elements, but it is sufficient to consider only one element of each of
the 767 non-trivial conjugacy classes.  Let
$\ket{b_0},\ldots,\ket{b_\ell-1}$ be a basis of a possibly degenerate
eigenspace of an element of $J_d$.  A generic vector in this
eigenspace has the form
\begin{equation}\label{eq:gen_state}
\ket{\phi_0}=\sum_{j=0}^{\ell-1}(x_{2j}+i x_{2j+1})\ket{b_j},
\end{equation}
where $x_0,\ldots,x_{2\ell-1}$ are real variables and $i^2=-1$.
Combining Eqs.~(\ref{eq:SICPOVM:state}), (\ref{eq:Heisenberg_orbit}),
and (\ref{eq:gen_state}) yields a system of polynomial equations for
the variables $x_j$.  Using the computer algebra system \Magma{}
\cite{magma}, we check whether this system has a solution.  While the
polynomial equations are stated over some cyclotomic field
$\Q(\omega_m)$, where $\omega_m:=\exp(2\pi i/m)$ is a primitive $m$-th
root of unity, we find a solution in the field
$\K:=\Q(\sqrt{3},\sqrt{7},\theta_1,\theta_2,\theta_3,i)$ which is an
algebraic extension of degree 96 over the rational numbers $\Q$. The
field $\K$ is generated by $\sqrt{3}$, $\sqrt{7}$, the complex number
$i$ with $i^2=-1$, and the real algebraic numbers $\theta_j$ given by
\begin{displaymath}
\theta_1:=\sqrt{\frac{3\sqrt{21}+9}{224}}, 
\qquad
\theta_3:=\sqrt{\frac{3\sqrt{21}+3\sqrt{7}+7\sqrt{3}+21}{148384}},
\end{displaymath}
and $\theta_2$ is the real root of the polynomial  
\begin{displaymath}
f(x)=x^3-\frac{\sqrt{21}-3}{28}x+\frac{21-5\sqrt{21}}{126}\theta_1.
\end{displaymath}
One solution for the initial state is
$\ket{\phi_0}=\theta_3(v_1,v_2,v_3,v_4,v_5,v_6)^t$, where the values
of $v_j$ are given in Fig.~\ref{fig:SICPOVM6}.  As mentioned before, a
global change of basis by an element of the Jacobi group yields a
SIC-POVM that is invariant under the Heisenberg group, too.  The
action of the Jacobi group $J_6$ on the Heisenberg group $H_6$ modulo
the center corresponds to the group $SL(2,\Z_6)$ of order $144$.  But
as the initial state $\ket{\phi_0}$ is fixed by an element of order
three, we get only $144/3=48$ different SIC-POVMs that are invariant
under the Heisenberg group.  Starting with the complex conjugated
state $\ket{\phi_0}$ (or replacing $\Pi_0$ by its transpose $\Pi_0^t$,
respectively) we get 48 different SIC-POVMs.  In total we obtain 96
SIC-POVMs that are invariant under the Heisenberg group, all related
by complex conjugation or by a global basis change corresponding to
conjugation by an element of $J_6$.  Note that this agrees with the
number of SIC-POVMs listed in \cite[Table 1]{RBKSC04}.

\begin{figure}[hbt]
$$
\begin{array}{r@{}c@{}l}
v_1&{}:={}&\Bigl(\bigl(336(\sqrt{7}-\sqrt{21})\theta_1-42\sqrt{21}-42\sqrt{3}-126\sqrt{7}-378\bigr)\theta_2^2\\
&&+\bigl(56(3\sqrt{7}-2\sqrt{3}+3)\theta_1+3\sqrt{21}-21\sqrt{3}+9\sqrt{7}+63\bigr)\theta_2\\
&&+(168-24\sqrt{21}-56\sqrt{3}+24\sqrt{7})\theta_1+6\sqrt{21}+18\sqrt{3}-6\sqrt{7}-6\Bigr)i\\
&&+\bigl(336(\sqrt{7}+\sqrt{21})\theta_1+42\sqrt{21}-42\sqrt{3}-126\sqrt{7}+378\bigr)\theta_2^2\\
&&+\bigl(56(3\sqrt{7}-2\sqrt{3}-3)\theta_1-3\sqrt{21}-21\sqrt{3}+9\sqrt{7}-63\bigr)\theta_2\\
&&+(24\sqrt{21}-56\sqrt{3}+24\sqrt{7}-168)\theta_1-6\sqrt{21}+18\sqrt{3}-6\sqrt{7}+6,\\
\noalign{\smallskip}
v_2&:=&\Bigl(\bigl(672(\sqrt{7}-\sqrt{21})\theta_1-168\sqrt{3}+504\bigr)\theta_2^2\\
&&+\bigl(28(3\sqrt{21}+5\sqrt{3}-3\sqrt{7}-15)\theta_1-42\sqrt{3}+126\bigl)\theta_2\\
&&+(336-48\sqrt{21}-112\sqrt{3}+48\sqrt{7})\theta_1-12\sqrt{21}-12\sqrt{3}+12\sqrt{7}+36\Bigr)i\\
&&-(84\sqrt{21}-252\sqrt{3}-252\sqrt{7}+252)\theta_2^2\\
&&+(84(\sqrt{21}+\sqrt{3}-3\sqrt{7}-1)\theta_1-6\sqrt{21}+18\sqrt{7})\theta_2-24\sqrt{3}+24,\\
\noalign{\smallskip}
v_3&:=&6(\sqrt{7}-\sqrt{3})i+6\sqrt{21}+12\sqrt{3}-12\sqrt{7}-18\\
\noalign{\smallskip}
v_4&:=&\Bigl(\bigl(336(\sqrt{7}-\sqrt{21})\theta_1+126\sqrt{21}-42\sqrt{3}-126\sqrt{7}+126\bigr)\theta_2^2\\
&&+\bigl(56(6-3\sqrt{21}-2\sqrt{3}+3\sqrt{7})\theta_1-9\sqrt{21}-21\sqrt{3}+9\sqrt{7}+63\bigr)\theta_2\\
&&+((168-24\sqrt{21}-56\sqrt{3}+24\sqrt{7})\theta_1+6\sqrt{21}+18\sqrt{3}-6\sqrt{7}-54)\Bigr)i\\
&&+\bigl(336(\sqrt{21}-3\sqrt{7})\theta_1+42\sqrt{21}-378\sqrt{3}-126\sqrt{7}+378\bigr)\theta_2^2\\
&&+\bigl(168(\sqrt{3}-1)\theta_1-3\sqrt{21}+63\sqrt{3}+9\sqrt{7}-63\bigr)\theta_2\\
&&+(24\sqrt{21}+168\sqrt{3}-72\sqrt{7}-168)\theta_1
+6-6\sqrt{21}-6\sqrt{3}+18\sqrt{7},\\
\noalign{\smallskip}
v_5&:=&\Bigl(\bigl(672\sqrt{7}\theta_1+84\sqrt{21}-168\sqrt{3}+252\bigr)\theta_2^2\\
&&-\bigl((84\sqrt{21}-140\sqrt{3}+84\sqrt{7}-84)\theta_1-6\sqrt{21}+42\sqrt{3}\bigr)\theta_2\\
&&-(112\sqrt{3}\theta_1-48\sqrt{7}\theta_1+12\sqrt{3}-12\sqrt{7}+24)\Bigr)i\\
&&+(672\sqrt{7}\theta_1-84\sqrt{21}-168\sqrt{3}-252)\theta_2^2\\
&&+\bigl((84\sqrt{21}+140\sqrt{3}-84\sqrt{7}-84)\theta_1-6\sqrt{21}-42\sqrt{3}\bigr)\theta_2\\
&&-112\sqrt{3}\theta_1+48\sqrt{7}\theta_1-12\sqrt{3}+12\sqrt{7}+24,\\
\noalign{\smallskip}
v_6&:=&6(\sqrt{7}-\sqrt{3})i-6\sqrt{21}+18.
\end{array}
$$
\fcaption{Explicit solution for the initial state
  $\ket{\phi_0}=\theta_3(v_1,v_2,v_3,v_4,v_5,v_6)^t$ of a SIC-POVM in
  dimension $d=6$.\label{fig:SICPOVM6}}
\end{figure}

\section{Mutually Unbiased Bases}
In the previous section we have considered generalized quantum
measurements with $d^2$ elements that allow the reconstruction of the
$d^2-1$ real parameters of a quantum state from the measurement
statistics.  If one allows only projective measurements, each
measurement provides only $d-1$ independent real parameters.  Hence we
need at least $d+1$ different orthonormal bases
$B_j:=\{\ket{\psi^j_k}\colon k=0,\ldots,d-1\}$ for the projective
measurements.  Again, the measurement results should be maximally
independent which leads to the following condition for the basis
states $\ket{\psi^j_k}$:
\begin{equation}\label{eq:MUB:state}
|\ip{\psi^j_k}{\psi^l_m}|^2 =\begin{cases}
1/d&\text{for $j\ne l$,}\\
\delta_{k,m} & \text{for $j=l$.}
\end{cases}
\end{equation}
We call a single vector $\ket{\psi}$ unbiased with respect to a set
$\mathcal{B}$ of vectors iff
\begin{equation}
|\ip{\psi}{\psi_k}|^2=1/d\qquad\mbox{for all $\ket{\psi_k}\in\mathcal{B}$.}
\end{equation}
The maximal size of a set of mutually unbiased bases in dimension $d$
is $d+1$.  Constructions for such maximal sets of MUBs are known when
the dimension $d$ is the power of a prime number (see
\cite{BBRV02,KlRo04} and references therein).  For non-prime-powers, a
lower bound is given by:
\begin{lemma}[see \cite{KlRo04}]
Let $N(d)$ denote the maximal size of a set of mutually unbiased bases
in dimension $d$.  Furthermore, let $d=p_1^{e_1}\cdot\ldots\cdot p_r^{e_r}$ be the
factorization of $d$ into powers of distinct primes $p_j$.  Then
\begin{equation}
N(d)\ge\min\{N(p_1^{e_1}),N(p_2^{e_2}),\ldots,N(p_r^{e_r})\}.
\end{equation}
In particular, $N(d)\ge 3$ for all $d\ge 2$.
\end{lemma}
In order to obtain three MUBs in any dimension, we use the following
construction of MUBs which is based on partitioning a set of unitary
matrices that are mutually orthogonal into subsets of commuting
matrices.
\begin{theorem}\label{theorem:UEB}
Let $\mathcal{B}=\mathcal{C}_1\cup\ldots\cup\mathcal{C}_\mu$ with $\mathcal{C}_j
\cap \mathcal{C}_l=\{I\}$ for $j\ne l$ be a set of $\mu(d-1)+1$ unitary
matrices which are mutually orthogonal with respect to the inner
product $\langle A,B\rangle :=\Tr(AB^\dagger)$.  Furthermore, let each
class $\mathcal{C}_j$ of the partition of $\mathcal{B}$ contain $d$
commuting matrices $U_{j,t}\in U(d,\C)$, $0\le t\le d-1$, where
$U_{j,0}:=I$.  For fixed $j$, let the basis $B_j$ contain the joint
eigenvectors $\ket{\psi^j_{k}}$ of the matrices $U_{j,t}$.  Then the
bases $B_j$ form a set of $\mu$ mutually unbiased bases, i.e.,
$$
|\ip{\psi^j_{k}}{\psi^l_{m}}|^2=1/d\qquad\mbox{for $j\ne l$.}
$$
\end{theorem}
\proof{
The statement follows directly from the proof of Theorem~3.2 in
\cite{BBRV02}.  
}

This theorem allows the construction of at least three MUBs in any dimension.
\begin{lemma}\label{lemma:MUB_XYZ}
For fixed $k$, $1\le k\le d-1$, $\gcd(k,d)=1$, the eigenvectors of the
operators $\{X, Z,XZ^k\}$ form three mutually unbiased bases. Using
the Fourier transformed basis, it follows that the eigenvectors of the
operators $\{X, Z,X^k Z\}$ form three mutually unbiased bases, too.
\end{lemma}
\proof{
Let $\mathcal{C}_A:=\{A^0,A,A^2,\ldots,A^{d-1}\}$ for $A=X,Z,XZ^k$.
Clearly, the matrices in each set $\mathcal{C}_A$ commute.  Modulo the
center of $H_d$, each set $\mathcal{C}_A$ can be written as
$\{X^{\lambda a} Z^{\lambda b}\colon \lambda=0,\ldots,d-1\}$ for fixed
$(a,b)\in \Z_d\times \Z_d$.  Hence the sets $\mathcal{C}_A$ correspond
to the $x$-axes, the $z$-axes, and a line with slope $k$.  For
$\gcd(k,d)=1$, the intersection of any two sets contains only the
identity matrix.  The second part of the statement follows from the
fact that the Fourier transform interchanges the operators $X$ and
$Z$, i.e., the Fourier transform corresponds to the interchange of the
coordinates of $\Z_d\times \Z_d$.
}

As far as we know, it is still open what the maximal size $N(6)$ of a
set of mutually unbiased bases in dimension six is.  Zauner
conjectured that $N(6)=3$ \cite{Zau99}.  This conjecture is supported
by the ---so far mere numerical--- similarity between the number of
MUBs and the number of mutually orthogonal latin squares
\cite{KlRo04}.  

From Lemma~\ref{lemma:MUB_XYZ}, we know how to construct three MUBs in
dimension six.  In order to find four MUBs, one can start with these
three bases and search for another basis that is mutually unbiased
with respect to the three initial bases.  A more general approach is
to start with only two of the bases, e.g., the eigenbases of $X$ and
$Z$.  But as we will show next, even then one can construct no more
than three mutually unbiased bases.
\begin{theorem}\label{theorem:MUB_XZ}
Let $B_X$ and $B_Z$ denote the eigenbasis of the operators $X$ and
$Z$, respectively.  Then there are exactly 48 normalized vectors
$\ket{v}\in\C^6$ with $v_1=1/\sqrt{6}$ such that
\begin{displaymath}
|\ip{w}{v}|^2=1/6\qquad\text{for all $\ket{w}\in B_X\cup B_Z$}.
\end{displaymath}
From the 48 vectors one can construct 16 different orthonormal bases
$B_i$ such that the three bases $\MUB_i:=\{B_X,B_Z,B_i\}$ are mutually
unbiased.  All $\MUB_i$ are maximal in the sense that there is no
vector $\ket{\psi}$ that is unbiased with respect to all three bases
in $\MUB_i$.
\end{theorem}
\proof{
Let $\ket{v}\in\C^6$ be an arbitrary vector that is unbiased with
respect to $B_X$ and $B_Z$.  The squared norm of the inner product of
$\ket{v}$ with an eigenvector of $Z$, i.e., an element of the standard
basis, must be equal to $1/6$.  Hence the squared norm of each
component of $\ket{v}$ is $1/6$.  Multiplying $\ket{v}$ by a phase
factor, one can, without loss of generality, assume that the first
component of $\ket{v}$ is $1/\sqrt{6}$.  Hence the vector $\ket{v}$
has the form
\begin{displaymath}
\ket{v}=\frac{1}{\sqrt{6}}(1,x_1+i x_6,x_2+i x_7,x_3+i x_8,x_4+i x_9,x_5+i x_{10})^t,
\end{displaymath}
where the variables $x_k$ are real and obey the equations
$x_k^2+x_{k+5}^2=1$.  From the inner product of $\ket{v}$ with the
eigenvectors of $X$ we get additional polynomial equations for the
variables $x_k$.  Using the computer algebra system \Magma{}
\cite{magma} one can show that this system of polynomial equations has
48 real solutions for the variables $x_k$.  The resulting set of
vectors $\mathcal{V}=\{\ket{v^\ell}\colon \ell=1,\ldots,48\}$ given in the
Appendix contains 16 orthonormal bases $B_i$, each of which is
unbiased with respect to $B_X$ and $B_Z$ (see
Fig.~\ref{fig:solutions_ZX} for the index sets of these bases).

Assume that we could find a vector that is unbiased with respect to
$B_X$, $B_Z$, and some $B_i$.  Hence $\mathcal{V}$ contains a vector that
is unbiased with respect to the six vectors in $B_i$. Each vector in
$\mathcal{V}$ is unbiased with respect to either 4 or 12 other vectors in
$\mathcal{V}$.  In Fig.~\ref{fig:unbiased_ZX} we list those vectors that
are unbiased with respect to 12 other vectors. None of these
12-element sets contains one of the bases $B_i$, thus all $\MUB_i$ are
maximal.}

\begin{figure}[hbt]
\small
\begin{displaymath}
\begin{array}{l@{}l@{}l}
\begin{array}[t]{l}
B_{1} :=\{1,5,9,14,18,22\}\\
B_{2} :=\{1,6,10,14,18,21\}\\
B_{3} :=\{2,5,9,13,17,22\}\\
B_{4} :=\{2,6,10,13,17,21\}\\
B_{5} :=\{3,8,12,16,20,23\}\\
B_{6} :=\{3,24,38,39,42,43\}
\end{array}&
\begin{array}[t]{l}
\phantom{B_{10}}\llap{$B_{7}$} :=\{4,7,11,15,19,24\}\\
\phantom{B_{10}}\llap{$B_{8}$} :=\{4,23,37,40,41,44\}\\
\phantom{B_{10}}\llap{$B_{9}$} :=\{7,20,26,27,30,31\}\\
B_{10}:=\{8,19,25,28,29,32\}\\
B_{11}:=\{11,16,33,35,46,48\}
\end{array}&
\begin{array}[t]{l}
B_{12}:=\{12,15,34,36,45,47\}\\
B_{13}:=\{25,31,33,40,42,47\}\\
B_{14}:=\{26,32,34,39,41,48\}\\
B_{15}:=\{27,29,36,38,44,46\}\\
B_{16}:=\{28,30,35,37,43,45\}
\end{array}
\end{array}
\end{displaymath}
\fcaption{The 16 bases that are unbiased with respect to the eigenbases $B_X$ and
  $B_Z$ of $X$ and $Z$.\label{fig:solutions_ZX}}.
\end{figure}

\begin{figure}[hbt]
\small
\begin{displaymath}
\begin{array}{ll}
\begin{array}{|r|l|}
\hline
&\multicolumn{1}{c|}{\text{unbiased vectors}}\\
\hline
 1 & \{ 15, 16, 19, 20, 26, 29, 35, 39, 40, 43, 44, 47 \}\\
 2 & \{ 15, 16, 19, 20, 25, 30, 36, 39, 40, 43, 44, 48 \}\\
 5 & \{ 11, 12, 23, 24, 27, 28, 31, 32, 33, 38, 41, 45 \}\\
 6 & \{ 11, 12, 23, 24, 27, 28, 31, 32, 34, 37, 42, 46 \}\\
 9 & \{  7,  8, 23, 24, 27, 32, 33, 34, 37, 42, 45, 46 \}\\
10 & \{  7,  8, 23, 24, 28, 31, 33, 34, 38, 41, 45, 46 \}\\
\hline
\end{array}&
\begin{array}{|r|l|}
\hline
&\multicolumn{1}{c|}{\text{unbiased vectors}}\\
\hline
13 & \{  3,  4, 19, 20, 26, 29, 35, 36, 40, 43, 47, 48 \}\\
14 & \{  3,  4, 19, 20, 25, 30, 35, 36, 39, 44, 47, 48 \}\\
17 & \{  3,  4, 15, 16, 25, 26, 29, 30, 35, 39, 44, 47 \}\\
18 & \{  3,  4, 15, 16, 25, 26, 29, 30, 36, 40, 43, 48 \}\\
21 & \{  7,  8, 11, 12, 27, 32, 33, 37, 38, 41, 42, 45 \}\\
22 & \{  7,  8, 11, 12, 28, 31, 34, 37, 38, 41, 42, 46 \}\\
\hline
\end{array}
\end{array}
\end{displaymath}
\fcaption{Vectors in $\mathcal{V}$ that are unbiased with respect to at
  least 6 other vectors.\label{fig:unbiased_ZX}}
\end{figure}
Using the automorphism group of the Heisenberg group, we can extend
this result  as follows:
\begin{corollary}\label{coro:MUB_YZ}
Theorem~\ref{theorem:MUB_XZ} still holds when replacing the eigenbases
$B_X$ and $B_Z$ of $X$ and $Z$, respectively, by the eigenbases of the
operators $X^aZ^b$ and $X^{a'}Z^{b'}$, where $ab'-a'b=1\bmod 6$.
\end{corollary}
\proof{
The action of the Jacobi group $J_d$ on the Heisenberg group $H_d$
modulo the center corresponds to the group $SL(2,\Z_d)$.  As
$ab'-a'b=1\bmod 6$, there exists a matrix $S\in J_6$ whose action on
$H_6$ corresponds to
$$
\begin{pmatrix}a&a'\\ b&b'\end{pmatrix},
$$
i.e. $X\entspricht(1,0)$ is mapped to $X^a Z^b\entspricht(a,b)$ and
$Z\entspricht(0,1)$ is mapped to $X^{a'} Z^{b'}\entspricht(a',b')$.
Hence the new bases can be obtained from $B_X$ and $B_Z$ via the
global change of basis $S$.
} 

In particular, this implies that in dimension six the MUBs of
Lemma~\ref{lemma:MUB_XYZ} with three bases are maximal, i.e., there is
no vector $\ket{v}\in\C^6$ such that $|\ip{w}{v}|^2=1/6$ for all
$\ket{w}\in B_X\cup B_{XZ^k}\cup B_Z$, $k=1,5$.

Finally, assume that we start with the eigenbases of two arbitrary
operators $A,B\in H_6$ that have non-degenerate eigenspaces.
Furthermore, let the intersection of the cyclic subgroups $H_A$ and
$H_B$ of $H_6$ generated by $A$ and $B$, respectively, be contained in
the center of $H_6$, i.e., $H_a\cap H_B\subseteq \zeta(H_6)$.  The two
subgroups correspond to lines in $\Z_6\times \Z_6$ that intersect only
in the origin.  From Theorem~\ref{theorem:UEB} it follows that the two
eigenbases bases are mutually unbiased.  As the group $SL(2,\Z_6)$
acts transitively on the pairs of lines with six points that intersect
only in the origin $(0,0)$, from Theorem~\ref{theorem:MUB_XZ} it
follows that the two bases cannot be contained in a set of four MUBs
in dimension six.  

Therefore, if a set of four or more MUBs in dimension six exists, the
bases cannot be related to the Heisenberg group.

\section{MUBs in Dimension 4}
As $4$ is a prime power, a maximal set with $5$ MUBs exists.  It can,
e.g., be constructed using the common eigenvectors of the following sets of
two-qubit Pauli matrices
\begin{alignat*}{8}
&\{id\otimes\sigma_x,\sigma_x\otimes id,\sigma_x\otimes\sigma_x\}, &\:
&\{\sigma_x\otimes\sigma_x,\sigma_y\otimes\sigma_y,\sigma_z\otimes\sigma_z\}, &\: 
&\{\sigma_x\otimes\sigma_y,\sigma_y\otimes\sigma_z,\sigma_z\otimes\sigma_x\},\\
&\{\sigma_x\otimes\sigma_z,\sigma_y\otimes\sigma_x,\sigma_z\otimes\sigma_y\}, &\: 
&\{id\otimes \sigma_z,\sigma_z\otimes id,\sigma_z\otimes\sigma_z\}.
\end{alignat*}
However, if we use the shift operator $X$ and the phase operator $Z$
as defined in (\ref{eq:Heisenberg}), we get again only three MUBs.
Direct computation using \Magma{} yields
\begin{proposition}\label{prop:dim4}
Starting with the eigenbases $B_Z$ and $B_X$ of the operators $Z$ and
$X$, respectively, given as the row-vectors of the following matrices
\begin{alignat*}{5}
B_Z:=
\begin{pmatrix}
1&0&0&0\\
0&1&0&0\\
0&0&1&0\\
0&0&0&1
\end{pmatrix}\qquad\text{and}\qquad
B_X:=\frac{1}{2}
\begin{pmatrix}
1& 1& 1& 1\\
1& i&-1&-i\\
1&-1& 1&-1\\
1&-i&-1& i
\end{pmatrix},
\end{alignat*}
the vectors of any third unbiased basis are given by the rows of the matrix
\begin{alignat*}{5}
B_3:=\frac{1}{2}
\begin{pmatrix}
1& e^{ia}& 1&-e^{ia}\\
1&-e^{ia}& 1& e^{ia}\\
1& e^{ib}&-1& e^{ib}\\
1&-e^{ib}&-1&-e^{ib}
\end{pmatrix}
\qquad\text{where $a,b\in[0,\pi)$}.
\end{alignat*}
For all choices of the parameters $a$ and $b$, the three MUBs are
maximal in the sense that there is no further vector that is unbiased
with respect to all three bases.
\end{proposition}

\section{Conclusions}
We have explicitly constructed a SIC-POVM in dimension six, the
smallest dimension that is not a prime power.  Unfortunately, we do
not know whether this implies that also a maximal set of MUBs---or
just more than three MUBs---in dimension six exist.  As there is no
affine plane of order six, we cannot make use of the geometric
analogies between SIC-POVMs and MUBs pointed out by Wootters.  Our
results rather support the conjecture that there are at most three
MUBs in dimension six.  This is true when starting with the eigenbasis
of some operators of the Heisenberg group with its geometric structure
over $\Z_d\times \Z_d$.

It seems as if the relation between MUBs and finite geometries was
stronger than the relation between SIC-POVMs and finite geometries.
Yet, it is still an open problem to determine the maximal number of
mutually unbiased bases in non-prime-power dimensions, even for the
smallest case $d=6$.

\nonumsection{Acknowledgements}
\noindent
The author acknowledges many interesting discussions with Martin
R{\"o}tteler, Betina Schnepf, and Robert Zeier.  In particular, he
would like to thank Thomas Decker for drawing his attention to the
problem of constructing SIC-POVMs and for many valuable comments in
the process of writing this paper.  Part of this work was supported
by Landesstiftung Baden-W\"urttem\-berg gGmbH (AZ~1.1322.01).

Proposition \ref{prop:dim4} is an answer to a question posed by
William Wootters while discussing the first version of this paper.
Finally, I'd like to thank Li Yu for pointing out an omission in
Lemma~\ref{lemma:MUB_XYZ}.

\nonumsection{References}

\nonumsection{Appendix}\label{anhang}
\noindent
In the following we list the $48$ vectors that are unbiased with
respect to the eigenbases $B_X$ and $B_Z$ of $X$ and $Z$.  Note that
the vectors are not normalized, the first coordinate is set to one.
Here $\omega=\exp(2\pi i/12)$ denotes a primitive $12$-th root of
unity and $\theta=\sqrt{(-2\omega^3 + 4\omega + 3)/48}$.

\begin{footnotesize}\raggedright
\def\w{\omega}
\def\r{\theta}
\def\c{,}
\catcode`,\active
\def,{\c\,\penalty5{}}
\begin{enumerate}
\item $\!\!\bigl(1,\w^5,1,-\w^3,-\w^2,-\w^3\bigr)$
\item $\!\!\bigl(1,\w^5,-\w^2,-\w^3,-\w^2,\w^5\bigr)$
\item $\!\!\bigl(1,\w^5,(2\w^2-6\w+2)\r+(2\w^3-\w^2-\w+1)/2,(-8\w^3+16\w-12)\r-\w^3+2\w^2-1,(12\w^3-8\w^2-12\w+16)\r+\w^3-\w^2+\w,(-4\w^3+6\w^2+2\w-6)\r+(\w^2-\w+1)/2\bigr)$
\item $\!\!\bigl(1,\w^5,(-2\w^2+6\w-2)\r+(2\w^3-\w^2-\w+1)/2,(8\w^3-16\w+12)\r-\w^3+2\w^2-1,(-12\w^3+8\w^2+12\w-16)\r+\w^3-\w^2+\w,(4\w^3-6\w^2-2\w+6)\r+(\w^2-\w+1)/2\bigr)$
\item $\!\!\bigl(1,-\w,1,\w^3,\w^4,\w^3\bigr)$
\item $\!\!\bigl(1,-\w,\w^4,\w^3,\w^4,-\w\bigr)$
\item $\!\!\bigl(1,-\w,(6\w^3-2\w^2-6\w+4)\r+(-\w^3+\w^2-\w)/2,(-8\w^3+16\w-12)\r+\w^3-2\w^2+1,(8\w^2-12\w+8)\r-2\w^3+\w^2+\w-1,(2\w^3-6\w^2+2\w)\r+(\w^3-\w^2-\w+2)/2\bigr)$
\item $\!\!\bigl(1,-\w,(-6\w^3+2\w^2+6\w-4)\r+(-\w^3+\w^2-\w)/2,(8\w^3-16\w+12)\r+\w^3-2\w^2+1,(-8\w^2+12\w-8)\r-2\w^3+\w^2+\w-1,(-2\w^3+6\w^2-2\w)\r+(\w^3-\w^2-\w+2)/2\bigr)$
\item $\!\!\bigl(1,\w^3,\w^4,\w^3,1,-\w\bigr)$
\item $\!\!\bigl(1,\w^3,-\w^2,\w^3,1,-\w^5\bigr)$
\item $\!\!\bigl(1,\w^3,(-6\w^3+4\w^2-2)\r+(-\w^3+2\w-1)/2,(-8\w^3+16\w-12)\r+\w^3-2\w^2+1,(12\w^3-16\w^2+8)\r+\w^3-2\w+1,(2\w^3-4\w+6)\r+(-\w^3+2\w^2-1)/2\bigr)$
\item $\!\!\bigl(1,\w^3,(6\w^3-4\w^2+2)\r+(-\w^3+2\w-1)/2,(8\w^3-16\w+12)\r+\w^3-2\w^2+1,(-12\w^3+16\w^2-8)\r+\w^3-2\w+1,(-2\w^3+4\w-6)\r+(-\w^3+2\w^2-1)/2\bigr)$
\item $\!\!\bigl(1,-\w^3,\w^4,-\w^3,1,\w\bigr)$
\item $\!\!\bigl(1,-\w^3,-\w^2,-\w^3,1,\w^5\bigr)$
\item $\!\!\bigl(1,-\w^3,(-6\w^3+4\w^2-2)\r+(-\w^3+2\w-1)/2,(8\w^3-16\w+12)\r-\w^3+2\w^2-1,(12\w^3-16\w^2+8)\r+\w^3-2\w+1,(-2\w^3+4\w-6)\r+(\w^3-2\w^2+1)/2\bigr)$
\item $\!\!\bigl(1,-\w^3,(6\w^3-4\w^2+2)\r+(-\w^3+2\w-1)/2,(-8\w^3+16\w-12)\r-\w^3+2\w^2-1,(-12\w^3+16\w^2-8)\r+\w^3-2\w+1,(2\w^3-4\w+6)\r+(\w^3-2\w^2+1)/2\bigr)$
\item $\!\!\bigl(1,\w,1,-\w^3,\w^4,-\w^3\bigr)$
\item $\!\!\bigl(1,\w,\w^4,-\w^3,\w^4,\w\bigr)$
\item $\!\!\bigl(1,\w,(6\w^3-2\w^2-6\w+4)\r+(-\w^3+\w^2-\w)/2,(8\w^3-16\w+12)\r-\w^3+2\w^2-1,(8\w^2-12\w+8)\r-2\w^3+\w^2+\w-1,(-2\w^3+6\w^2-2\w)\r+(-\w^3+\w^2+\w-2)/2\bigr)$
\item $\!\!\bigl(1,\w,(-6\w^3+2\w^2+6\w-4)\r+(-\w^3+\w^2-\w)/2,(-8\w^3+16\w-12)\r-\w^3+2\w^2-1,(-8\w^2+12\w-8)\r-2\w^3+\w^2+\w-1,(2\w^3-6\w^2+2\w)\r+(-\w^3+\w^2+\w-2)/2\bigr)$
\item $\!\!\bigl(1,-\w^5,1,\w^3,-\w^2,\w^3\bigr)$
\item $\!\!\bigl(1,-\w^5,-\w^2,\w^3,-\w^2,-\w^5\bigr)$
\item $\!\!\bigl(1,-\w^5,(2\w^2-6\w+2)\r+(2\w^3-\w^2-\w+1)/2,(8\w^3-16\w+12)\r+\w^3-2\w^2+1,(12\w^3-8\w^2-12\w+16)\r+\w^3-\w^2+\w,(4\w^3-6\w^2-2\w+6)\r+(-\w^2+\w-1)/2\bigr)$
\item $\!\!\bigl(1,-\w^5,(-2\w^2+6\w-2)\r+(2\w^3-\w^2-\w+1)/2,(-8\w^3+16\w-12)\r+\w^3-2\w^2+1,(-12\w^3+8\w^2+12\w-16)\r+\w^3-\w^2+\w,(-4\w^3+6\w^2+2\w-6)\r+(-\w^2+\w-1)/2\bigr)$
\item $\!\!\bigl(1,(4\w^3-6\w^2-2\w+6)\r+(\w^2-\w+1)/2,(6\w^3-2\w^2-6\w+4)\r+(-\w^3+\w^2-\w)/2,-\w^3,(2\w^2-6\w+2)\r+(2\w^3-\w^2-\w+1)/2,(2\w^3-6\w^2+2\w)\r+(-\w^3+\w^2+\w-2)/2\bigr)$
\item $\!\!\bigl(1,(4\w^3-6\w^2-2\w+6)\r+(\w^2-\w+1)/2,(-12\w^3+8\w^2+12\w-16)\r+\w^3-\w^2+\w,(8\w^3-16\w+12)\r-\w^3+2\w^2-1,(-2\w^2+6\w-2)\r+(2\w^3-\w^2-\w+1)/2,\w^5\bigr)$
\item $\!\!\bigl(1,(4\w^3-6\w^2-2\w+6)\r+(-\w^2+\w-1)/2,(12\w^3-8\w^2-12\w+16)\r+\w^3-\w^2+\w,(8\w^3-16\w+12)\r+\w^3-2\w^2+1,(2\w^2-6\w+2)\r+(2\w^3-\w^2-\w+1)/2,-\w^5\bigr)$
\item $\!\!\bigl(1,(4\w^3-6\w^2-2\w+6)\r+(-\w^2+\w-1)/2,(-6\w^3+2\w^2+6\w-4)\r+(-\w^3+\w^2-\w)/2,\w^3,(-2\w^2+6\w-2)\r+(2\w^3-\w^2-\w+1)/2,(2\w^3-6\w^2+2\w)\r+(\w^3-\w^2-\w+2)/2\bigr)$
\item $\!\!\bigl(1,(-4\w^3+6\w^2+2\w-6)\r+(\w^2-\w+1)/2,(12\w^3-8\w^2-12\w+16)\r+\w^3-\w^2+\w,(-8\w^3+16\w-12)\r-\w^3+2\w^2-1,(2\w^2-6\w+2)\r+(2\w^3-\w^2-\w+1)/2,\w^5\bigr)$
\item $\!\!\bigl(1,(-4\w^3+6\w^2+2\w-6)\r+(\w^2-\w+1)/2,(-6\w^3+2\w^2+6\w-4)\r+(-\w^3+\w^2-\w)/2,-\w^3,(-2\w^2+6\w-2)\r+(2\w^3-\w^2-\w+1)/2,(-2\w^3+6\w^2-2\w)\r+(-\w^3+\w^2+\w-2)/2\bigr)$
\item $\!\!\bigl(1,(-4\w^3+6\w^2+2\w-6)\r+(-\w^2+\w-1)/2,(6\w^3-2\w^2-6\w+4)\r+(-\w^3+\w^2-\w)/2,\w^3,(2\w^2-6\w+2)\r+(2\w^3-\w^2-\w+1)/2,(-2\w^3+6\w^2-2\w)\r+(\w^3-\w^2-\w+2)/2\bigr)$
\item $\!\!\bigl(1,(-4\w^3+6\w^2+2\w-6)\r+(-\w^2+\w-1)/2,(-12\w^3+8\w^2+12\w-16)\r+\w^3-\w^2+\w,(-8\w^3+16\w-12)\r+\w^3-2\w^2+1,(-2\w^2+6\w-2)\r+(2\w^3-\w^2-\w+1)/2,-\w^5\bigr)$
\item $\!\!\bigl(1,(2\w^3-4\w+6)\r+(-\w^3+2\w^2-1)/2,(-6\w^3+4\w^2-2)\r+(-\w^3+2\w-1)/2,\w^3,(6\w^3-4\w^2+2)\r+(-\w^3+2\w-1)/2,(-2\w^3+4\w-6)\r+(-\w^3+2\w^2-1)/2\bigr)$
\item $\!\!\bigl(1,(2\w^3-4\w+6)\r+(-\w^3+2\w^2-1)/2,(12\w^3-16\w^2+8)\r+\w^3-2\w+1,(-8\w^3+16\w-12)\r+\w^3-2\w^2+1,(-6\w^3+4\w^2-2)\r+(-\w^3+2\w-1)/2,\w^3\bigr)$
\item $\!\!\bigl(1,(2\w^3-4\w+6)\r+(\w^3-2\w^2+1)/2,(6\w^3-4\w^2+2)\r+(-\w^3+2\w-1)/2,-\w^3,(-6\w^3+4\w^2-2)\r+(-\w^3+2\w-1)/2,(-2\w^3+4\w-6)\r+(\w^3-2\w^2+1)/2\bigr)$
\item $\!\!\bigl(1,(2\w^3-4\w+6)\r+(\w^3-2\w^2+1)/2,(-12\w^3+16\w^2-8)\r+\w^3-2\w+1,(-8\w^3+16\w-12)\r-\w^3+2\w^2-1,(6\w^3-4\w^2+2)\r+(-\w^3+2\w-1)/2,-\w^3\bigr)$
\item $\!\!\bigl(1,(-2\w^3+6\w^2-2\w)\r+(\w^3-\w^2-\w+2)/2,(2\w^2-6\w+2)\r+(2\w^3-\w^2-\w+1)/2,\w^3,(6\w^3-2\w^2-6\w+4)\r+(-\w^3+\w^2-\w)/2,(-4\w^3+6\w^2+2\w-6)\r+(-\w^2+\w-1)/2\bigr)$
\item $\!\!\bigl(1,(-2\w^3+6\w^2-2\w)\r+(\w^3-\w^2-\w+2)/2,(-8\w^2+12\w-8)\r-2\w^3+\w^2+\w-1,(8\w^3-16\w+12)\r+\w^3-2\w^2+1,(-6\w^3+2\w^2+6\w-4)\r+(-\w^3+\w^2-\w)/2,-\w\bigr)$
\item $\!\!\bigl(1,(-2\w^3+6\w^2-2\w)\r+(-\w^3+\w^2+\w-2)/2,(8\w^2-12\w+8)\r-2\w^3+\w^2+\w-1,(8\w^3-16\w+12)\r-\w^3+2\w^2-1,(6\w^3-2\w^2-6\w+4)\r+(-\w^3+\w^2-\w)/2,\w\bigr)$
\item $\!\!\bigl(1,(-2\w^3+6\w^2-2\w)\r+(-\w^3+\w^2+\w-2)/2,(-2\w^2+6\w-2)\r+(2\w^3-\w^2-\w+1)/2,-\w^3,(-6\w^3+2\w^2+6\w-4)\r+(-\w^3+\w^2-\w)/2,(-4\w^3+6\w^2+2\w-6)\r+(\w^2-\w+1)/2\bigr)$
\item $\!\!\bigl(1,(2\w^3-6\w^2+2\w)\r+(\w^3-\w^2-\w+2)/2,(8\w^2-12\w+8)\r-2\w^3+\w^2+\w-1,(-8\w^3+16\w-12)\r+\w^3-2\w^2+1,(6\w^3-2\w^2-6\w+4)\r+(-\w^3+\w^2-\w)/2,-\w\bigr)$
\item $\!\!\bigl(1,(2\w^3-6\w^2+2\w)\r+(\w^3-\w^2-\w+2)/2,(-2\w^2+6\w-2)\r+(2\w^3-\w^2-\w+1)/2,\w^3,(-6\w^3+2\w^2+6\w-4)\r+(-\w^3+\w^2-\w)/2,(4\w^3-6\w^2-2\w+6)\r+(-\w^2+\w-1)/2\bigr)$
\item $\!\!\bigl(1,(2\w^3-6\w^2+2\w)\r+(-\w^3+\w^2+\w-2)/2,(2\w^2-6\w+2)\r+(2\w^3-\w^2-\w+1)/2,-\w^3,(6\w^3-2\w^2-6\w+4)\r+(-\w^3+\w^2-\w)/2,(4\w^3-6\w^2-2\w+6)\r+(\w^2-\w+1)/2\bigr)$
\item $\!\!\bigl(1,(2\w^3-6\w^2+2\w)\r+(-\w^3+\w^2+\w-2)/2,(-8\w^2+12\w-8)\r-2\w^3+\w^2+\w-1,(-8\w^3+16\w-12)\r-\w^3+2\w^2-1,(-6\w^3+2\w^2+6\w-4)\r+(-\w^3+\w^2-\w)/2,\w\bigr)$
\item $\!\!\bigl(1,(-2\w^3+4\w-6)\r+(-\w^3+2\w^2-1)/2,(6\w^3-4\w^2+2)\r+(-\w^3+2\w-1)/2,\w^3,(-6\w^3+4\w^2-2)\r+(-\w^3+2\w-1)/2,(2\w^3-4\w+6)\r+(-\w^3+2\w^2-1)/2\bigr)$
\item $\!\!\bigl(1,(-2\w^3+4\w-6)\r+(-\w^3+2\w^2-1)/2,(-12\w^3+16\w^2-8)\r+\w^3-2\w+1,(8\w^3-16\w+12)\r+\w^3-2\w^2+1,(6\w^3-4\w^2+2)\r+(-\w^3+2\w-1)/2,\w^3\bigr)$
\item $\!\!\bigl(1,(-2\w^3+4\w-6)\r+(\w^3-2\w^2+1)/2,(-6\w^3+4\w^2-2)\r+(-\w^3+2\w-1)/2,-\w^3,(6\w^3-4\w^2+2)\r+(-\w^3+2\w-1)/2,(2\w^3-4\w+6)\r+(\w^3-2\w^2+1)/2\bigr)$
\item $\!\!\bigl(1,(-2\w^3+4\w-6)\r+(\w^3-2\w^2+1)/2,(12\w^3-16\w^2+8)\r+\w^3-2\w+1,(8\w^3-16\w+12)\r-\w^3+2\w^2-1,(-6\w^3+4\w^2-2)\r+(-\w^3+2\w-1)/2,-\w^3\bigr)$
\end{enumerate}
\end{footnotesize}


\begin{thebibliography}{10}

\bibitem{BBRV02}
{\sc S.~Bandyopadhyay, P.~O. Boykin, V.~Roychowdhury, and F.~Vatan}, {\em {A
  New Proof for the Existence of Mutually Unbiased Bases}}, Algorithmica, 34
  (2002), pp.~512--528.
\newblock Preprint quant-ph/0103162.

\bibitem{BeSc98}
{\sc R.~Berndt and R.~Schmidt}, {\em {Elements of the Representation Theory of
  the Jacobi Group}}, vol.~163 of Progress in mathematics, Birkh{\"a}user,
  Basel, 1998.

\bibitem{magma}
{\sc W.~Bosma, J.~J. Cannon, and C.~Playoust}, {\em {The Magma Algebra System
  I: The User Language}}, Journal of Symbolic Computation, 24 (1997),
  pp.~235--266.

\bibitem{GHW04}
{\sc K.~S. Gibbons, M.~J. Hoffman, and W.~K. Wootters}, {\em {Discrete Phase
  Space Based on Finite Fields}}. Physical Review A, 70 (2004), 062101.
\newblock Preprint quant-ph/0401155, 2004.

\bibitem{KlRo04}
{\sc A.~Klappenecker and M.~R{\"o}tteler}, {\em {Constructions of Mutually
  Unbiased Bases}}, in {Finite Fields and Applications: 7th International
  Conference, Fq7}, G.~L. Mullen, A.~Poli, and H.~Stichtenoth, eds.,
  Lecture Notes in Computer Science, vol. 2984, Springer, 2004,
  pp.~137--144. 
\newblock Preprint quant-ph/0309120.

\bibitem{RBKSC04}
{\sc J.~M. Renes, R.~Blume-Kohout, A.~J. Scott, and C.~M. Caves}, {\em
  {Symmetric Informationally Complete Quantum Measurements}}, Journal of
  Mathematical Physics, 45 (2004), pp.~2171--2180.
\newblock Preprint quant-ph/0310075.

\bibitem{Sco09}
{\sc A.~Scott}, 
\url{http://www.cit.gu.edu.au/~ascott/sicpovms/}, accessed 2009-05-25.

\bibitem{Tarry1900}
{\sc G.~Tarry}, {\em {Le Probl{\`e}me de $36$ Officiers}}, Comptes Rendues de
  l'Assoc. Fran{\c{c}}ais Avanc. Sci. Naturel, 1 (1900), pp.~122--123.

\bibitem{Weyl28}
{\sc H.~Weyl}, {\em Gruppentheorie und Quantenmechanik}, Hirzel, Leipzig, 1928.

\bibitem{Weyl50}
\leavevmode\vrule height 2pt depth -1.6pt width 23pt, {\em {The Theory of
  Groups and Quantum Mechanics}}, Dover Publications, New York, 1950.
\newblock Translated from the 2nd rev. German ed. by H.~P.~Robertson.

\bibitem{Woo04:fin_geom}
{\sc W.~K. Wootters}, {\em {Quantum Measurements and Finite
    Geometry}}, {\em Foundations of Physics}, 36 (2006), pp.~112--126. 
\newblock Preprint quant-ph/0406032v2, 2004.

\bibitem{Zau99}
{\sc G.~Zauner}, {\em {Quantendesigns -- Grundz{\"u}ge einer nichtkommutativen
  Designtheorie}}.
\newblock Dissertation, Universit{\"at} Wien, 1999.

\end{thebibliography}
\end{document}